\documentclass[prx,aps,amsfonts,floatfix,superscriptaddress,longbibliography,twocolumn]{revtex4-2}

\usepackage{graphicx}
\usepackage{dcolumn}
\usepackage{bm}
\usepackage{mathbbol}

\usepackage{amsmath,amssymb,amsthm}
\usepackage{times}
\usepackage{graphicx}
\usepackage{bm}
\usepackage{color}
\usepackage{multi row} 

\usepackage[mathscr]{euscript}
\usepackage[utf8]{inputenc} 

\begin{document}

\title{Speck of Chaos}

\author{Lea F. Santos}
\affiliation{Department of Physics, Yeshiva University, New York, New York 10016, USA}
\author{Francisco P\'erez-Bernal}
\address{Dep. CC. Integradas y Centro de Estudios Avanzados en F\'isica,
Matem\'aticas y Computaci\'on. Fac. CC. Experimentales, Universidad de Huelva, Huelva 21071, \& Instituto Carlos I de F\'isica Te\'orica y Computacional, Universidad de Granada, Granada 18071, Spain}
\author{E. Jonathan Torres-Herrera}
\address{Instituto de F\'isica, Benem\'erita Universidad Aut\'onoma de Puebla,
Apt. Postal J-48, Puebla, 72570, Mexico}

\begin{abstract}
It has been shown that, despite being local, a perturbation applied to a single site of the one-dimensional XXZ model is enough to bring this interacting integrable spin-1/2 system to the chaotic regime. Here, we show that this is not unique to this model, but happens also to the Ising model in a transverse field and to the spin-1 Lai-Sutherland chain. The larger the system is, the smaller the amplitude of the local perturbation for the onset of chaos. We focus on two indicators of chaos, the correlation hole, which is a dynamical tool, and the distribution of off-diagonal elements of local observables, which is used in the eigenstate thermalization hypothesis. Both methods avoid spectrum unfolding and can detect chaos even when the eigenvalues are not separated by symmetry sectors.
\end{abstract}

\maketitle

\section{Introduction}
The term quantum chaos, as used in this work, refers to properties of the spectrum and eigenstates that are similar to those found in full random matrices, such as strongly correlated eigenvalues~\cite{HaakeBook,Guhr1998,StockmannBook} and eigenstates close to random vectors~\cite{Chirikov1985,Casati1993,Flambaum1994,Zelevinsky1996,Santos2012PRE,Borgonovi2016,Zyczkowski1990}. Level statistics as in random matrices are found also in some integrable models, but they are caused by finite-size effects~\cite{Sorathia2012,Torres2019} or change abruptly upon tiny variations of the Hamiltonian parameters~\cite{Benet2003,Relano2004}. Other definitions of quantum chaos include the short-time exponential growth of out-of-time order correlators~\cite{Maldacena2016PRD,Rozenbaum2017,Hashimoto2017,Garcia2018,Jalabert2018,Chavez2019} and diffusive transport~\cite{Jin2015,Iadecola2019,BertiniARXIV}, although exponential behaviors of four-point correlation functions appear also near critical points of integrable models~\cite{Pappalardi2018,Hummel2019,Pilatowsky2019,Xu2020,HashimotoARXIV} and ballistic transport has been observed in the chaotic single-defect XXZ model~\cite{Brenes2018}.

The one-dimensional clean spin-1/2 XXZ model represents an interacting integrable system, whose transport behavior has been extensively studied~\cite{Zotos1999,BertiniARXIV}. In 2004, it was shown that the model becomes chaotic if a single site has a Zeeman splitting different from that on the other sites~\cite{Santos2004} (see also~\cite{Santos2011,Gubin2012}). At first, it was thought that the transport behavior of this single-defect XXZ model was diffusive~\cite{Barisic2009}, but it was later concluded that it is ballistic~\cite{Brenes2018}.  In spite of that, the model shows all the expected properties of chaotic many-body quantum systems. As the system size increases, level repulsion and chaotic eigenstates emerge for smaller defect amplitudes~\cite{Torres2014PRE}, local observables satisfy the diagonal eigenstate thermalization hypothesis (ETH) \cite{Torres2014PRE,TorresKollmar2015}, and the system's long-time dynamics manifest spectral correlations~\cite{Torres2017Philo}.

The single-defect XXZ model has motivated studies of transport behavior in single-defect noninteracting models~\cite{Jansen2019} and searches for minimal chaotic models with electron-phonon coupling~\cite{Ljubotina2019}, but it was not until the beginning of 2020 that the model saw a significant resurgence of interest. It has since been employed in studies of many-body quantum chaos~\cite{Corps2020,PandeyARXIV}, thermalization~\cite{BrenesARXIVa,RichterARXIV}, quantum transport~\cite{BrenesARXIVb,ZnidaricARXIV}, and entanglement~\cite{Ashouri2020}. In the present work, we show that the onset of chaos due to a local onsite perturbation is not unique to the XXZ model. This is illustrated for the spin-1/2 Ising model in a transverse field and the spin-1 Lai-Sutherland chain. The former is among the simplest quantum systems that exhibit a critical point and, contrary to the XXZ model, it is solved without the Bethe Ansatz technique. The latter, which has a SU(3) symmetry, has been investigated in the context of Haldane gapped materials~\cite{Batchelor2004a,Batchelor2004b} and its transport behavior is receiving increasing attention~\cite{Dupont2020}.

What are the best ways to detect quantum chaos? The analysis of level statistics is the most common approach when one has direct access to the spectrum, as in nuclear physics~\cite{Guhr1998}. It requires the separation of the eigenvalues by symmetry sectors and, depending on the chaos indicator, also the unfolding of the spectrum. Other methods that have been put forward include the analysis of the structure of the eigenstates~\cite{Zelevinsky1996,Borgonovi2016,Chirikov1985,Casati1993,Flambaum1994,Santos2012PRE,Zyczkowski1990} and the entanglement entropy~\cite{Vidmar2017b,Vidmar2017}.  In Ref.~\cite{Beugeling2015} (see also~\cite{BrenesARXIVb,LeBlond2019,Khaymovich2019,Haque2019}), the distinction between integrable and chaotic models is based on the distribution of the off-diagonal matrix elements of local observables in each subspace. In Refs.~\cite{PandeyARXIV,VillazonARXIV}, a new chaos indicator based on the rate of deformations of the eigenstates under small perturbations bypasses the need to unfold the spectrum and to separate it by symmetries. Identifying all symmetries of a model is not always trivial, so having a way to detect chaos despite their presence is important in studies of both chaotic and integrable models. 

To leave no doubts about the chaotic nature of our single-defect models, we consider three indicators of chaos: level statistics, matrix elements of local observables, and also the correlation hole. We show that the distribution of the off-diagonal elements of local observables diagnoses chaos also when the energy levels are not separated by subspaces. However, eigenvalues, eigenstates, and matrix elements of observables are not easily accessible to experiments that focus on time evolutions, such as those with cold atoms and ion traps. Therefore, we promote the use of the correlation hole~\cite{Leviandier1986,Pique1987,Guhr1990,Wilkie1991,Hartmann1991,Delon1991,Lombardi1993,Alhassid1992,Michaille1999,Leyvraz2013,Torres2017Philo,Torres2018,Schiulaz2019,Lerma2019,CruzARXIV}, which is a dynamical tool to capture level repulsion and spectrum rigidity. This chaos indicator does not require unfolding the spectrum or separating it by symmetries~\cite{CruzARXIV}.  We discuss how the time scale for the onset of the correlation hole in the three single-defect models -- XXZ, Ising, and Lai-Sutherland chains -- depends on the defect amplitude and on the system size.

\section{Models} The Hamiltonians for the spin-1/2 XXZ model, spin-1/2 Ising model in a transverse field, and spin-1 Lai-Sutherland model~\cite{Uimin1970,Lai1974,Sutherland1975} in the presence of a single defect of amplitude $d$ in the middle of the chain are  respectively given by
\begin{equation}
H_{\text{XXZ}} = d J S_{L/2}^z + J\sum_{k=1}^{L-1} \left( S_k^x S_{k+1}^x + S_k^y S_{k+1}^y +\Delta S_k^z S_{k+1}^z \right)  ,
\label{ham1} 
\end{equation}
\begin{equation}
H_{\text{ZZ}} = d J S_{L/2}^z +   J h_{x} \sum_{k=1}^{L}  S_{k}^x - J\sum_{k=1}^{L-1}  S_k^z S_{k+1}^z  ,
\label{ham2} 
\end{equation}
\begin{eqnarray}
H_{\text{S1}} &=& d J S^z_{L/2} + J \sum_{k=1}^{L-1}  \left( S_k^x S_{k+1}^x +S_k^y S_{k+1}^y + S_k^z S_{k+1}^z \right) 
\nonumber \\
&+& J \sum_{k=1}^{L-1}  \left[  \left(S_k^x S_{k+1}^x \right)^2 + \left(S_k^y S_{k+1}^y \right)^2 + \left(S_k^z S_{k+1}^z \right)^2 \right]. \hspace{0.5 cm}
\label{ham3}
\end{eqnarray}
Above, $\hbar =1$, $L$ is the number of sites, $S^{x,y,z}_k $ are spin operators acting on site $k$,  $J$ is the coupling constant that sets the energy scale, $\Delta$ is the anisotropy of the XXZ model, and $h_x$ is the amplitude of the transverse field in the Ising model. Note that, contrary to the case of spin 1/2, the quadratic term in Eq.~(\ref{ham3}) is necessary to guarantee integrability when $d=0$.

Open boundary conditions are considered to avoid translational symmetry. To avoid parity and spin reversal, we add to $H_{\text{XXZ}}$ (\ref{ham1}) and $H_{\text{ZZ}}$ (\ref{ham2}) small impurities at the edges of the chain, $\epsilon_{1,L} J S_{1,L}^z$, where $\epsilon_{1,L}$ are random numbers in $[-0.1,0.1]$.  In the case of the spin-1 model, we add to $H_{\text{S1}}$ (\ref{ham3}) the term $\epsilon_1 J S_{1}^x$, which connects symmetry sectors where the total magnetization in the $z$-direction differs by $1$.  While for the spin-1/2 models, the onset of chaos requires placing the defect $d$ out of the borders~\cite{Santos2004}, for the spin-1 model, chaos emerges when the defect $d$ is on any site, including the edges. 

The parameters used are $\Delta=0.48$ and $h_x=0.84$. The XXZ model conserves total spin in the $z$-direction, so we study the largest subspace of dimension ${\cal D}_\text{XXZ}=L!/(L/2)!^2$. For the other two models:  ${\cal D}_\text{ZZ}=2^L$ and ${\cal D}_\text{S1}=3^L$.

\section{Level Statistics} The most used signature of quantum chaos is the distribution of spacings between nearest unfolded energy levels~\cite{noteUnfold}. For chaotic systems with real and symmetric Hamiltonian matrices, as the full random matrices from the Gaussian orthogonal ensemble (GOE), the level spacing distribution follows the Wigner-Dyson distribution~\cite{MehtaBook,Guhr1998}, $P_{\rm WD}(s) = (\pi s/2) \exp \left( -\pi s^2/4 \right)$, which indicates that the eigenvalues are highly correlated and repel each other. In integrable models, where the energy levels are uncorrelated and not prohibited from crossing, the level spacing distribution is usually Poissonian, $P_{\rm P}(s) = e^{-s}$, but exceptions include ``picket-fence''-kind of spectra~\cite{Berry1977,Pandey1991,Chirikov1995} and systems with an excessive number of degeneracies~\cite{Zangara2013}.

The crossover from integrability to chaos can be studied with an indicator that quantifies how close the level spacing distribution is to $P_{\rm WD}(s)$. An example is the value of $\beta$ obtained by fitting $P(s)$ with the Brody distribution~\cite{Brody1981} (see also~\cite{Izrailev1990}),
\begin{equation}
P_\beta (s) = (\beta+1) b s^{\beta} \exp(-b s^{\beta+1}), \hspace{0.4 cm} 
b =\left[ \Gamma \left( \frac{\beta+2}{\beta+1} \right) \right]^{\beta+1}
\label{Eq:beta}
\end{equation}
Chaotic systems give $\beta \sim 1$, while the Poissonian distribution leads to $\beta \sim 0$. 

\begin{figure}[ht!]
\includegraphics*[width=3.4in]{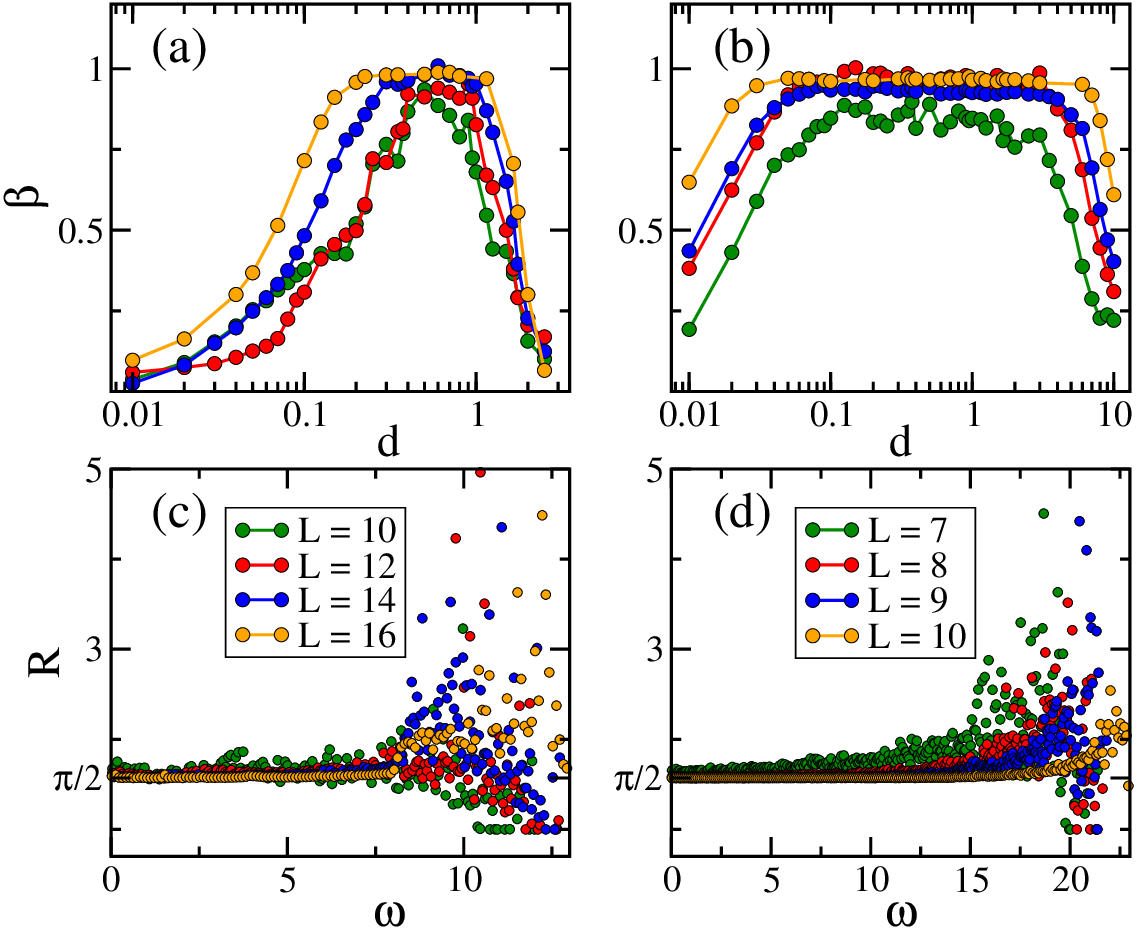}
\caption{Chaos indicator $\beta$ for various values of $d$ for the Ising model (a) and the spin-1 model (b) averaged over 20 realizations of border defects. The ratio $R$ from Eq.~(\ref{Eq:R}) vs $\omega$ for the Ising model (c) and spin-1 model (d) in the chaotic regime, $d=0.8$.  Following~\cite{BrenesARXIVb}, we consider only the eigenstates for which $(E_{\beta} +E_{\alpha})/2$ is in the interval $[\bar{E}-0.05(E_{\text{max}}-E_{\text{min}})/2, \bar{E}+0.05(E_{\text{max}}-E_{\text{min}})/2]$, where $\bar{E}$ is the middle of the spectrum and $E_{\text{max}}$ ($E_{\text{min}}$) is the largest (smallest) eigenvalue; bin width $d\omega=0.05$. }
\label{Fig01}
\end{figure}

In Figs.~\ref{Fig01} (a) and (b), we show $\beta$ as a function of the defect amplitude for the Ising (a) and the  Lai-Sutherland model (b). One sees that the range of values of $d$ for which $\beta \sim 1$ increases with system size, eliminating any suspicion that the appearance of the Wigner-Dyson distribution might have been a finite-size effect. A discussion about how the amplitude of the defect for the onset of chaos decreases as the system size grows is provided in Appendix~\ref{AppA} using for the that the single-defect XXZ model.

The level spacing distribution and the ratio of consecutive levels~\cite{Oganesyan2007,Atas2013} detect short-range correlations. A more complete analysis of level statistics calls for the study of long-range correlations as well, as measured, for example, with the level number variance~\cite{Guhr1998}. We verified that the level number variance for the three single-defect models with $d \sim 1$ approaches the GOE result as $L$ increases (not shown). 

An advantage of the ratio of consecutive levels over the level spacing distribution and the level number variance is that the ratio does not require unfolding the spectrum. However, a prerequisite for all three quantities is the separation of the eigenvalues by symmetry sectors. If we mix eigenvalues from different subspaces, Poissonian statistics may emerge even when the system is chaotic~\cite{Santos2009}.

\section{Eigenstate thermalization hypothesis} Indicators of ETH based on observables can also be used to detect quantum chaos without spectrum unfolding. In chaotic systems, the infinite-time averages of local observables approach thermodynamic averages as the system size increases. This is referred to as the diagonal ETH and has been verified for the single-defect XXZ model in~\cite{Torres2014PRE,TorresKollmar2015} and recently in~\cite{BrenesARXIVa}. In chaotic systems, the distribution of the off-diagonal matrix elements of local operators is Gaussian~\cite{Beugeling2015,LeBlond2019}, which is called the off-diagonal ETH and has been confirmed for the single-defect XXZ model as well~\cite{BrenesARXIVb}.

We studied the shape of the distribution of the off-diagonal elements of the operator that breaks the integrability of the Ising and  Lai-Sutherland models, that is, the distribution of $\langle \psi_{\beta}|S^z_{L/2}|\psi_{\alpha}\rangle $, where $|\psi_{\alpha,\beta}\rangle$ are the eigenstates of $H_{\text{ZZ}}$ (\ref{ham2})  and  $H_{\text{S1}}$ (\ref{ham3}). To confirm that the distribution is indeed Gaussian and therefore complies with ETH, one performs tests of normality, such as skewness and kurtosis. In Figs.~\ref{Fig01} (c) and (d), we show the results for the ratio~\cite{LeBlond2019},
\begin{equation}
R(\omega) = \overline{|\langle \psi_{\alpha}|S^z_{L/2}|\psi_{\beta}\rangle|^2}/\overline{|\langle \psi_{\alpha}|S^z_{L/2}|\psi_{\beta}\rangle|}^2,
\label{Eq:R}
\end{equation}
where the bar indicates the average over the off-diagonal elements for which the energy difference $\omega=|E_{\beta} -E_{\alpha}|$ lies in one of the bins of width $d\omega=0.05$. For a Gaussian distribution~\cite{Geary1935}, $R(\omega)=\pi/2$.

Figures~\ref{Fig01} (c) and (d) show that the range of values of $\omega$ for which $R(\omega)\sim \pi/2$ increases as the system size grows. This picture does not hold for the single-defect models in the integrable limit, where distributions other than Gaussian are found.

\section{Correlation Hole}
The results above substantiate that the single-defect models are chaotic. But which dynamical manifestation of chaos do they exhibit and how does it depend on the defect amplitude and system size? To answer these questions, we study the mean survival probability,
\begin{equation}
\langle \mathscr{S}_p(t)  \rangle \!\!=\!\!  \langle \left|\left<\Psi (0)|\Psi (t)\right>\right|^2  \rangle 
\!\!=\!\! \left\langle
 \sum_{\alpha,\beta=1}^{\cal D}  |C_{\alpha}^0|^2 |C_{\beta}^0|^2 e^{-i (E_{\alpha}-E_{\beta}) t} \!  \right\rangle
\label{eq:SP}
\end{equation}
where $C_{\alpha}^0= \left< \psi_{\alpha} |\Psi (0) \right>$ and $ \left< . \right>$ indicates an average over initial states that have energy $E_0=\left<\Psi (0)|H|\Psi (0)\right>$ close to the middle of the spectrum~\cite{noteAve}. The average is needed, because this quantity is not self-averaging at any time scale~\cite{Schiulaz2020}. The initial states used are eigenstates of the $z$-terms in $H$~(\ref{ham1})-(\ref{ham3}). The mean survival probability is related to the spectral form factor, $K(t) = \sum_{\alpha,\beta=1}^{\cal D} \langle e^{-i (E_{\alpha}-E_{\beta}) t } \rangle$.

The initial decay of the survival probability is determined by the shape and bounds of the energy distribution of the initial state~\cite{Torres2014PRA,Torres2014NJP,Tavora2016,Tavora2017}. The presence of correlated eigenvalues gets explicitly manifested later, when the dynamics resolve the discreteness of the spectrum and the mean survival probability develops a dip below its saturation point, known as correlation hole~\cite{Leviandier1986,Pique1987,Guhr1990,Wilkie1991,Hartmann1991,Delon1991,Lombardi1993,Alhassid1992,Michaille1999,Leyvraz2013,Torres2017Philo,Torres2018,Schiulaz2019,Lerma2019,CruzARXIV}, which appears also for experimental local observables~\cite{Torres2018,Schiulaz2019}. The use of the correlation hole as an alternative to detect level repulsion was first proposed for molecules with poor line resolution~\cite{Leviandier1986}.  The interval $t_{\text{m}} \leq t \leq t_{\text{H}}$, where the correlation hole is found, is limited by the point of its minimum, $t_{\text{m}}$,  and by the longest time scale of the system, the so-called Heisenberg time, $t_{\text{H}}$, which is inversely proportional to the mean level spacing. 

\begin{figure}[ht!]
\includegraphics*[width=3.4in]{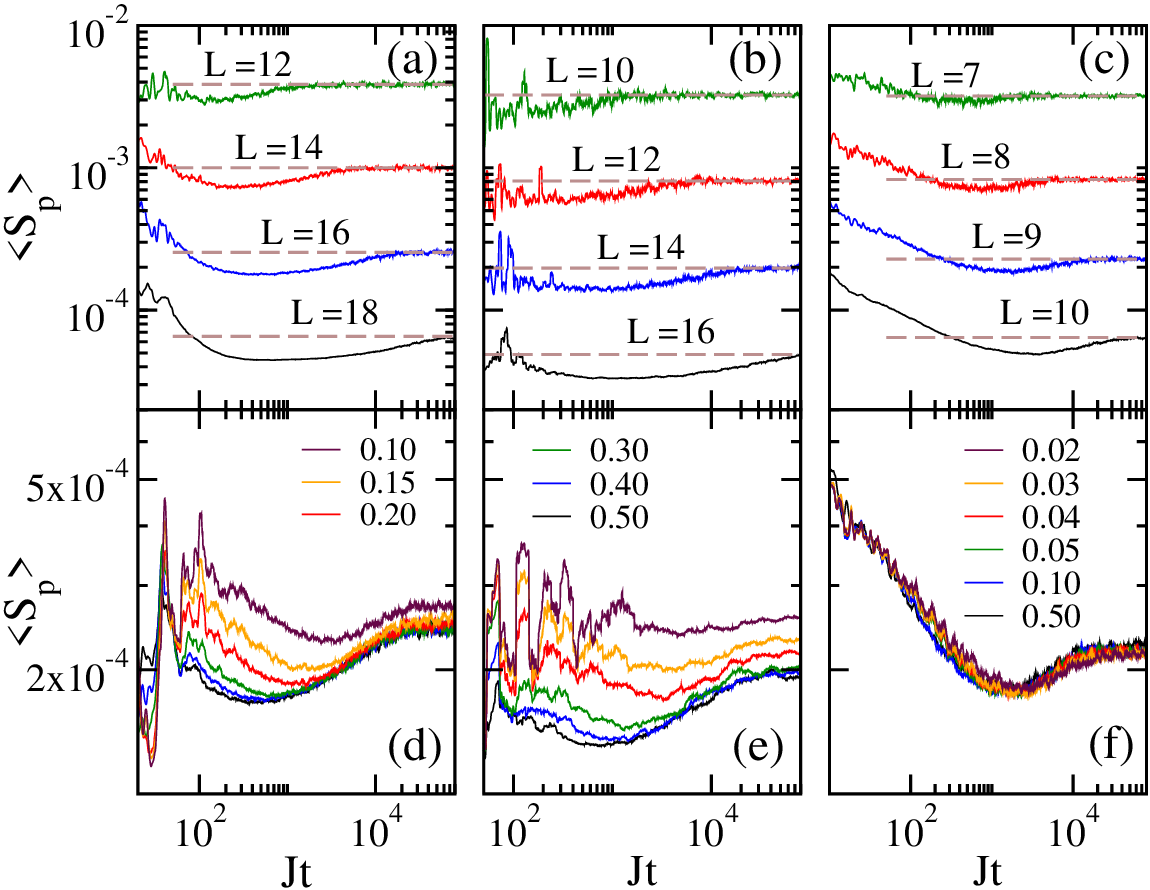}
\caption{Mean survival probability for the XXZ (a), (d), Ising (b), (e), and spin-1 (c), (f) model for different system sizes (a)-(c) and different defect amplitudes (d)-(f). Top row: $d=0.8$. Bottom row: $L=16$ (d), $L=14$ (e), and $L=9$ (f).}
\label{Fig02}
\end{figure}

The onset of the correlation hole for the single-defect models is analyzed in Fig.~\ref{Fig02}. The mean survival probability for the XXZ model is shown in Figs.~\ref{Fig02}~(a) and (d), for the Ising model in Figs.~\ref{Fig02}~(b) and (e), and for the spin-1 model in Figs.~\ref{Fig02}~(c) and (f). Curves for the systems at strong chaos ($d=0.8$) and for different systems sizes are displayed in Figs.~\ref{Fig02}~(a)-(c). They make it evident that $t_{\text{m}}$  grows exponentially with $L$ for the three cases. These long times are unrelated with the fact that the perturbation is local. They reflect instead the locality of the spin-spin couplings, which causes the gradual and slow spread of the initial many-body states in the Hilbert space. The correlation hole also gets elongated as the system size increases, since the growth constant of the exponential behavior of $t_{\text{H}}$ with $L$ is larger than that for $t_{\text{m}}$. These features are all very similar to those observed in chaotic systems with global perturbations and local couplings~\cite{Schiulaz2019}.

In Figs.~\ref{Fig02}~(d)-(f), we fix the system size and examine how $t_{\text{m}}$ depends on the defect amplitude. As $d$ decreases from 0.5 toward the integrable point, the correlation hole gets postponed to later times for the XXZ and the Ising model (see plots for $t_\text{m}$ vs $d$ in Appendix~\ref{AppB}). This is expected, since the approach to integrability reduces the correlations between the eigenvalues and the first ones to be eliminated are the long-range correlations. It calls attention, however, that the spin-1 model does not show the same behavior. In this case, as $d$ decreases below 0.5, the correlation hole is not displaced and even its depth is hardly affected [Fig.~\ref{Fig02}~(f)]. This raises the suspicion that the border defect $\epsilon_1 S^x_1$ may suffice to bring the Lai-Sutherland chain to the chaotic regime even when $d=0$. The reason why we did not notice this in  Fig.~\ref{Fig01}~(b) may be an indication that not all symmetries of this model were identified.

\section{Symmetries}  
\begin{figure}[t!]
\includegraphics*[width=3.2in]{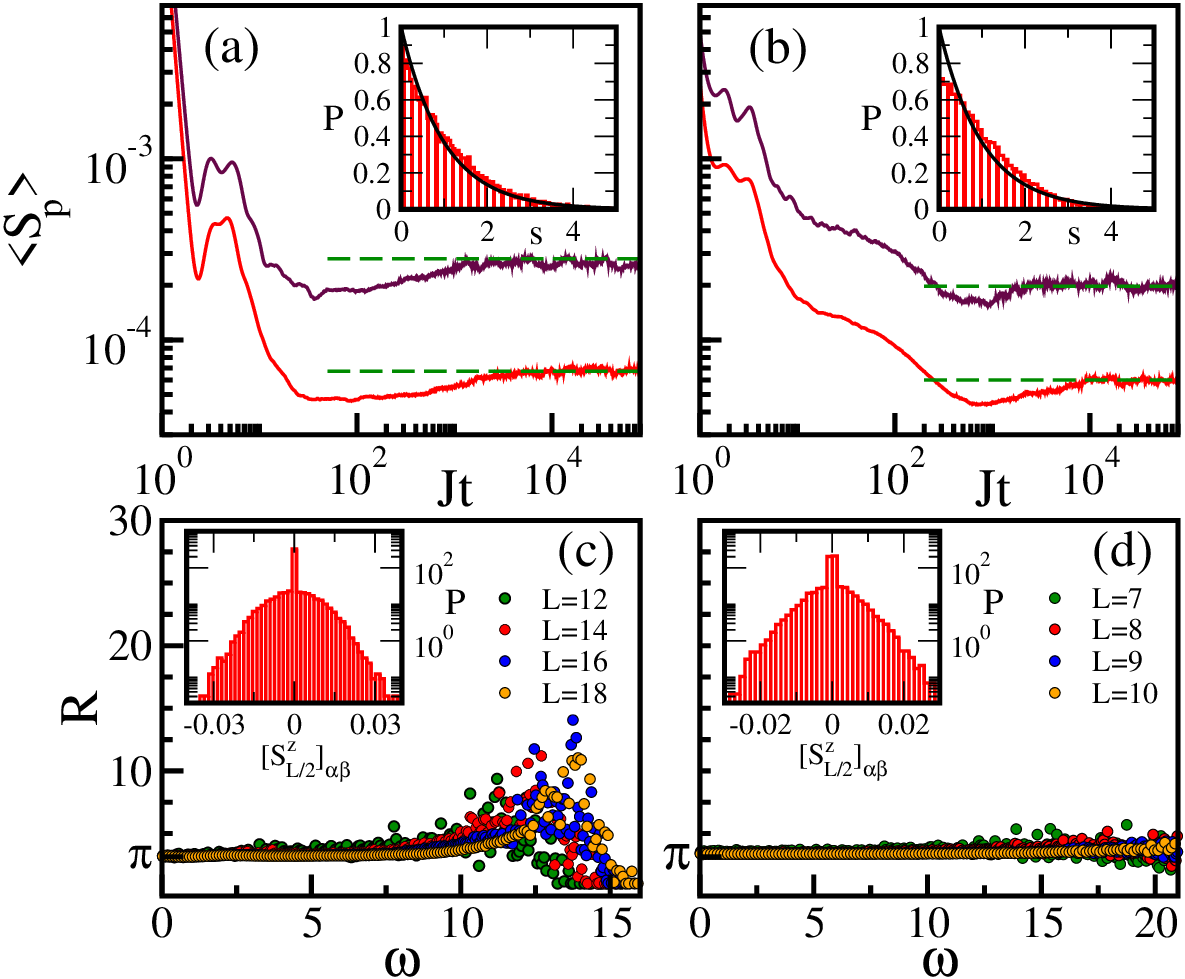}
\caption{Mean survival probability for the chaotic NNN spin-1/2 model [Eq.~(\ref{ham4})] with $L=16, 18$ (a) and for the spin-1 model with $L=9, 10$, $d=0$, and $\epsilon_1= 0.05$ (b). The horizontal line marks the saturation point. The insets show the corresponding level spacing distributions, $L=14$ (a) and $L=8$ (b). The ratio $R(\omega)$ is shown for the NNN model (c) and for the spin-1 model (d) for different system sizes. The insets show the distributions of $\langle \psi_{\alpha}|S^z_{L/2}|\psi_{\beta}\rangle$ for 200 eigenstates in the middle of the spectrum, $L=16$  (c) and $L=9$ (d). In (a) and (c): $\sum_k S^z_k=0$.
}
\label{Fig03}
\end{figure}

Poissonian level statistics may emerge in chaotic systems if the eigenvalues from different subspaces are mixed. This contrasts with the correlation hole, whose appearance requires only the presence of correlated eigenvalues, not their separation by symmetry sectors~\cite{CruzARXIV}. To illustrate this, we show in Fig.~\ref{Fig03}~(a) the mean survival probability for a spin-1/2 model that is known to be chaotic. It has couplings between nearest- and next-nearest neighbors (NNN) and is described by the following Hamiltonian,
\begin{equation}
H_{NNN} = J\sum_{k=1}^{L-1} \vec{S}_k \cdot \vec{S}_{k+1} +
0.9 J\sum_{k=1}^{L-2} \vec{S}_k \cdot \vec{S}_{k+2} .
\label{ham4}
\end{equation}
This system conserves total magnetization in the $z$-direction, total spin, and it also exhibits parity and spin reversal. The level spacing distribution in the inset of Fig.~\ref{Fig03}~(a) disregards these symmetries, apart from the $z$-magnetization, which results in a Poissonian distribution. The correlation hole, on the other hand, is evident in the main panel.

Similar results are found in Fig.~\ref{Fig03}~(b) for the spin-1 model with $d=0$ and a very small border defect, $\epsilon_1= 0.05$. (For comparison, see $\langle  \mathscr{S}_p(t) \rangle$ for the clean integrable spin-1 model in Appendix~\ref{AppC}.) The inset in Fig.~\ref{Fig03}~(b) shows a nearly Poissonian distribution, while the correlation hole is apparent in the main panel. This makes clear the power of the correlation hole as a dynamical tool to identify chaotic systems, but it suggests also its usefulness in the search for integrable models. Verifying whether a Hamiltonian, which may have unknown symmetries, remains integrable after small modifications, as done here, is a very hard problem that is often avoided. In studies of integrability,  the usual strategy is instead to build integrable Hamiltonians using for example the quantum Yang-Baxter equations. 

A natural question that arises from the discussions above is what happens to the off-diagonal ETH in the presence of symmetries~\cite{noteETH}. It turns out that it can still detect chaos, but $R(\omega)$ is no longer $\pi/2$. Chaos is now revealed by the flatness of the curves for $R(\omega)$ at values close to integer multiples of $\pi/2$, as seen in Figs.~\ref{Fig03}~(c)  for the NNN model and in Fig.~\ref{Fig03}~(d) for the spin-1 model. The specific value of $R(\omega)$ depends on the observable and the number of symmetry sectors. The following picture provides a simplified explanation. 

Suppose that one has two subspaces, each with $N={\cal D}/2$ different chaotic eigenstates, and the operator associated with the symmetry sectors commutes with the observable $O$ used to compute $R(\omega)$. The distribution of the $[N(N-1)]/2$ values of $\langle \psi_{\alpha}|O|\psi_{\beta}\rangle$ within each subspace is Gaussian, but by mixing the sectors, one also has $N^2$ values $\langle \psi_{\alpha}|O|\psi_{\beta}\rangle=0$ for the cases where  $|\psi_{\alpha}\rangle$ and $|\psi_{\beta}\rangle$ belong to different subspaces. As a result, $R(\omega)=\pi$, and the distribution of the off-diagonal elements is zero-inflated, similar to the ones seen in the insets of Figs.~\ref{Fig03}~(c) and (d). If instead of 2, one has $m$ subspaces, then $R(\omega)=m\pi/2$.

\section{Conclusions} In integrable quantum systems with many interacting particles, a local perturbation applied to a single site can be enough for the onset of chaos. Such small changes to a Hamiltonian calls for indicators that can detect chaos in the presence of symmetries, such as the correlation hole and the distribution of off-diagonal elements of local observables. These two methods combined may assist the identification of subspaces and the search for integrable models.

\begin{acknowledgments}
LFS was supported by the NSF grant No. DMR-1936006 and thanks Angela Foerster for discussions. This work has received funding from the Spanish National Research, Development, and Innovation plan (RDI plan) under the project PID2019-104002GB-C21. It has been partially supported by the Consejer\'{\i}a de Conocimiento, Investigaci\'on y Universidad, Junta de Andaluc\'{\i}a and European Regional Development Fund (ERDF), ref. SOMM17/6105/UGR and UHU-1262561. Computing resources were provided by the CEAFMC and Universidad de Huelva High Performance Computer (HPC@UHU) located in the Campus Universitario el Carmen and funded by FEDER/MINECO project UNHU-15CE-2848. E.J.T.-H. was funded from VIEP-BUAP (Grant Nos. MEBJ-EXC19-G, LUAGEXC19-G), Mexico. He is also grateful to LNS-BUAP for their supercomputing facility.
\end{acknowledgments}

\appendix

\section{Dependence on system size}
\label{AppA}
The curves in Figs.~1~(a)-(b) clearly indicate that as the system size increases, the amplitude of the defect needed for the onset of chaos decreases. But how exactly does $d$ decrease with $L$? A proper scaling analysis is hard when dealing with many-body systems and their exponentially large Hilbert spaces, since the numerical data are restricted to few system sizes. The analysis of level statistics requires the full exact diagonalization of the Hamiltonian matrices, which restricts the dimension of the Hilbert space to ${\cal D} \sim 10^5$. 

For the Ising model [Fig.~1~(c)], we have four points only and $L=10$ is too small to be used. For the spin-1 model [Fig.~1~(d)], in addition to having very few sizes available, we still need to identify and take into account the remaining symmetry (or symmetries) that we detected with the studies of the correlation hole. But before giving up the idea completely, let us have a closer look at the XXZ model.

\begin{figure}[ht!]
\includegraphics*[width=3.4in]{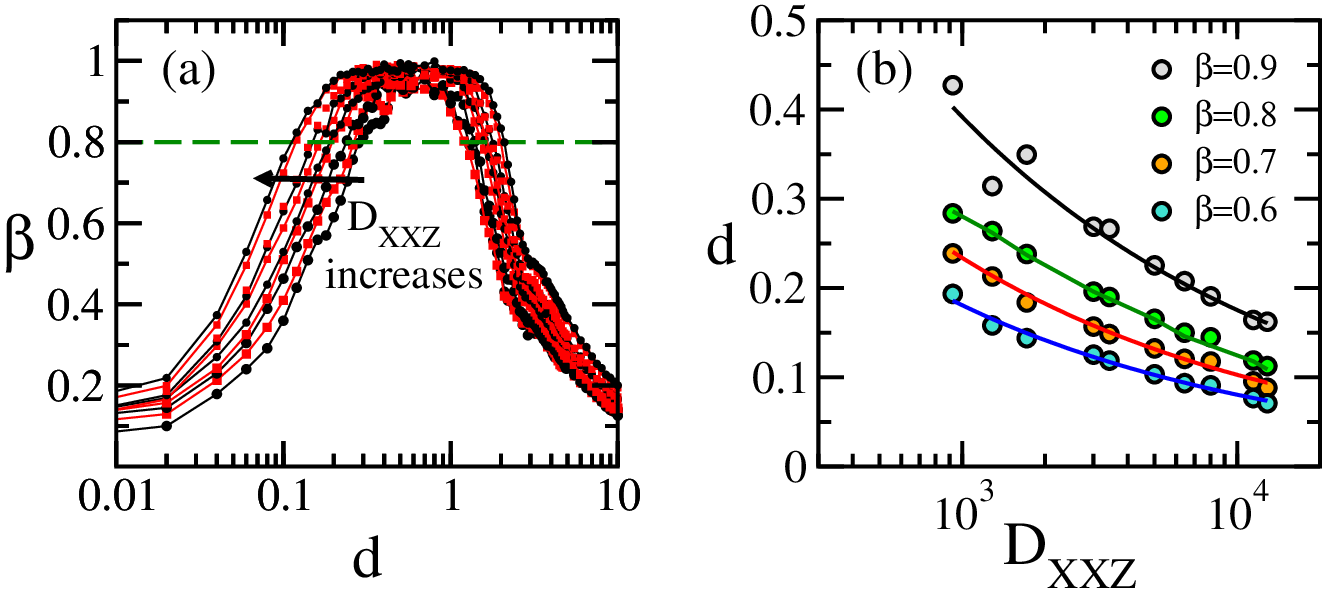}
\caption{Chaos indicator $\beta$ for various values of $d$ for the XXZ model (a) and value of $d$ as a function of the $\sum_k S^z_k$ subspace dimension ${\cal D}_{\text{XXZ}}$ for the four values of $\beta$ indicated (b). The values of $\{ L,n, {\cal D}_{\text{XXZ}} \}$ from the smallest dimension to the largest are: $\{ 12, 6, 924\}$, $\{ 13, 5, 1287\}$, $\{ 13, 6, 1716\}$, $\{ 14, 6, 3003\}$, $\{ 14, 7, 3432\}$, $\{ 15, 6, 5005\}$, $\{ 15, 7, 6435\}$,  $\{ 16, 7, 11440\}$, and $\{ 16, 8, 12870\}$. In (b): circles are numerical data and the fittings are indicated with the solid line: $d =2.04/{\cal D}_{\text{XXZ}}^{0.351} $ for $\beta=0.6$, $d =2.75/{\cal D}_{\text{XXZ}}^{0.357} $ for $\beta=0.7$, $d =3.24/{\cal D}_{\text{XXZ}}^{0.350} $ for $\beta=0.8$, and $d =4.37/{\cal D}_{\text{XXZ}}^{0.349} $ for $\beta=0.9$.
}
\label{Fig01SM}
\end{figure}

Contrary to the Ising and the Lai-Sutherland systems, the XXZ model conserves the total magnetization in the $z$-direction. One can then take advantage of this symmetry to get a larger number of points for the scaling analysis, that is, for the same system size $L$, we consider different values of $\sum_k S^z_k$ and therefore different values of ${\cal D}_{\text{XXZ}}=L!/[n! (L-n)!]$, where $n$ is the number of excitations. Naturally, since we are interested in many-body quantum chaos, we stay away from the dilute limit and consider only $n= L/2-1$ and $n=L/2$. In Fig.~\ref{Fig01SM}~(a), we show $\beta$ as a function of $d$ for nine values of ${\cal D}_{\text{XXZ}}$. The figure follows the same trend of Fig.~1~(c) and Fig.~1~(d), but now with more curves.

We select a threshold value for $\beta$ for which reasonable Wigner-Dyson distributions are seen and show in Fig.~\ref{Fig01SM}~(b) how the defect amplitude $d$ for the chosen $\beta$ decreases as ${\cal D}_{\text{XXZ}}$ increases. The decay of $d$ is evident for the four chosen values $\beta=0.6, 0.7, 0.8, 0.9$ and our studies suggest that $d \propto  {\cal D}_{\text{XXZ}}^{-0.35} $, but we cannot predict what may happen for larger systems sizes and cannot preclude an eventual halt on the decay of $d$.

\section{Dependence on defect amplitude}
\label{AppB}

In Figs.~2~(d)-(e), we fix the system size and observe that the time to reach the minimum of the correlation hole increases as the defect size decreases from $d= 0.5$ toward the integrable point. The hole also gets shallower and it should eventually disappear altogether when the integrable point has uncorrelated eigenvalues.

\begin{figure}[ht!]
\includegraphics*[width=3.3in]{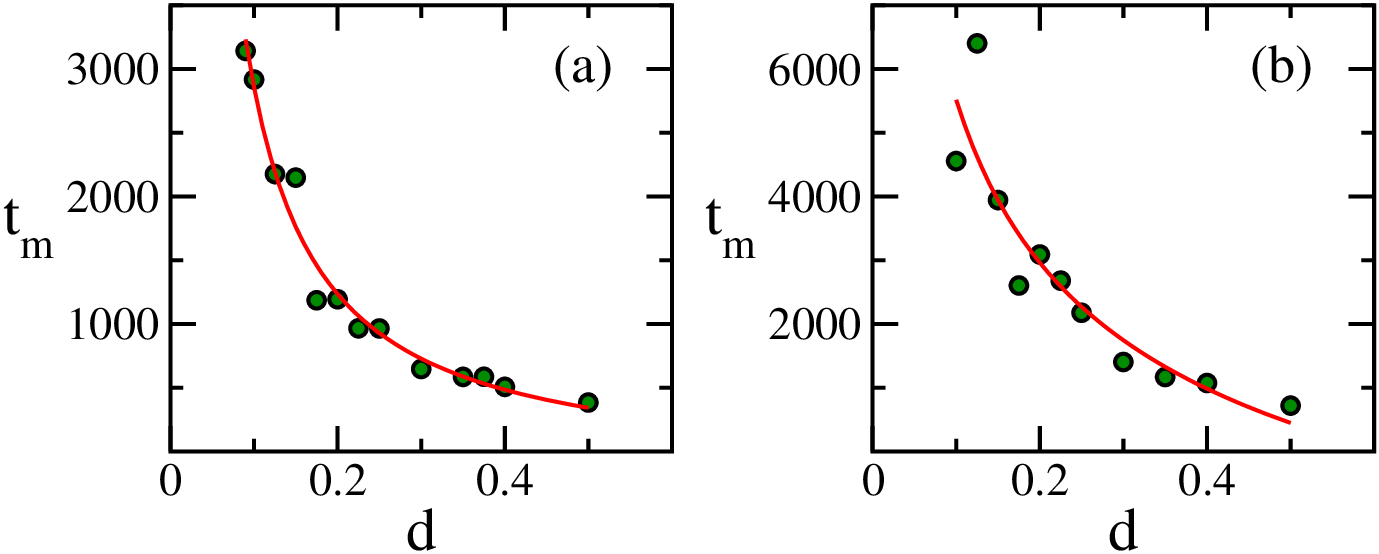}
\caption{Time for the minimum of the hole as a function of the defect amplitude for the XXZ model with $L=16$ (a) and for the Ising model with $L=14$ (b). Solid lines indicate: $t_{\text{m} }\propto d^{-1.1}$ in (a) and $t_{\text{m}} \propto d^{-0.4}$ in (b). 
}
\label{Fig02SM}
\end{figure}

In Fig.~\ref{Fig02SM}, we show how $t_{\text{m}}$ increases as $d$ decreases for the XXZ model (a) and the Ising model (b). The fitting gives $t_{\text{m} }\propto d^{-1.1}$ for the XXZ model and $t_{\text{m}} \propto d^{-0.4}$ for the Ising model. Similarly to the discussion in Sec.~I, these results prompt the question of how these behaviors depend on the system size. This is again hard to answer, due to the limitations in system sizes, but the correlation hole offers the great advantage of being a dynamical quantity. Methods other than exact diagonalization exist to study the time evolution of systems larger than those available to exact diagonalization and these techniques are constantly being improved. We may therefore presume that with the correlation hole we might be able to perform better scaling analysis than what we can now do with the eigenvalues. 

\section{Survival probability in the clean spin-1 model}
\label{AppC}

\begin{figure}[htb]
\includegraphics*[width=2.in]{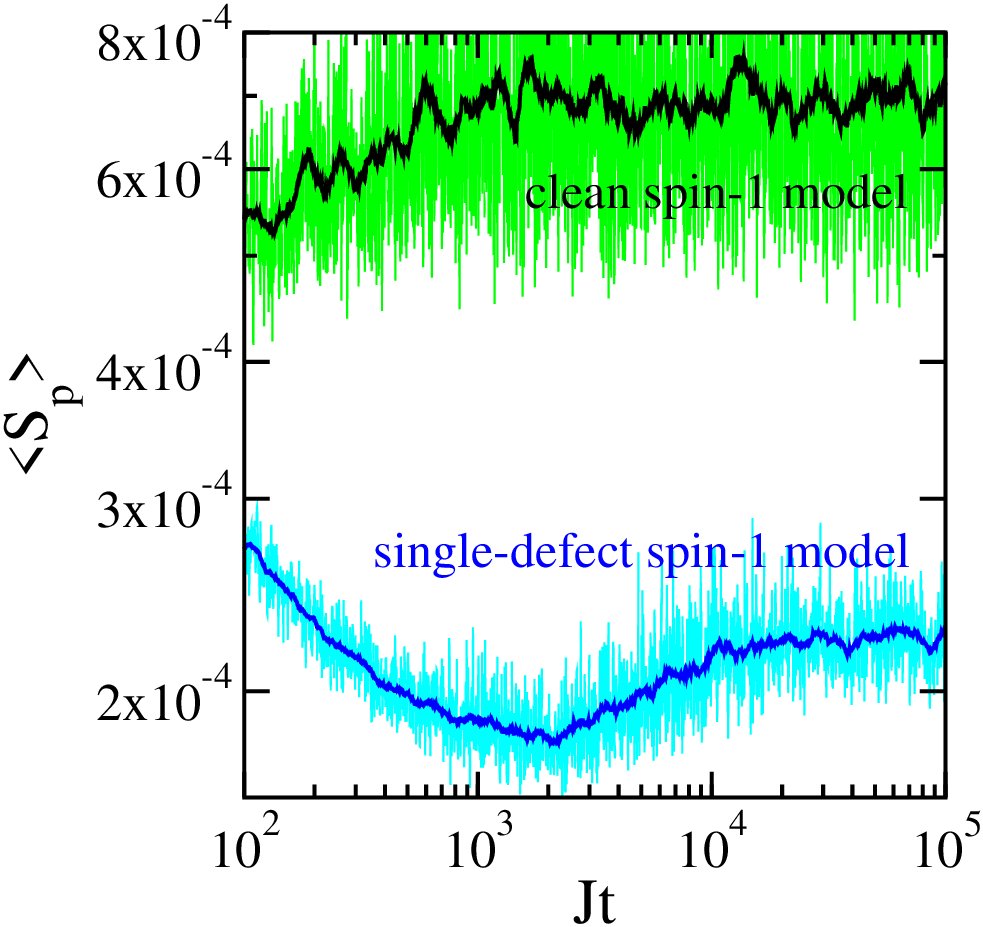}
\caption{Mean survival probability for the clean integrable spin-1 model ($d=\epsilon_1=0$) and for the chaotic single-defect spin-1 model ($d=0.5$ and $\epsilon_1 \in [-0.1,0.1]$) for $L = 9$. Dark solid lines represent running averages.
}
\label{Fig03SM}
\end{figure}

The mean survival probability for the integrable clean Lai-Sutherland spin-1 model with $d=\epsilon_1=0$ is shown in Fig.~\ref{Fig03SM} together with the curve for $d=0.5$ and $\epsilon_1=0$ to allow for the comparison between the two and with the case where $d=0$ and $\epsilon_1=0.05$ presented in the Fig.~3~(b). The same average over initial states with energies close to the mean of the spectrum is considered here also. For the clean case, there is no sign of the correlation hole caused by level repulsion of the kind seen in random matrices. At the integrable point, the spin-1 model has more symmetries, which results in the large fluctuations seen in Fig.~\ref{Fig03SM} even after a running average. The symmetries also bring the saturation value of $\langle \mathscr{S}_p(t) \rangle$ to a level higher than for the chaotic single-defect spin-1 model, as expected.

We note that correlations of other kinds, such as those caused by the Shnirelmann's peak observed in some integrable models, may lead to correlation holes of different forms and at different time scales~\cite{Torres2019}, but they are easily distinguished by the holes generated by quantum chaos, which can be described using random matrix theory~\cite{Schiulaz2019}.


\begin{thebibliography}{93}
\bibitem [{not({\natexlab{a}})}]{noteUnfold}%
  \BibitemOpen
  \href@noop {} {}\bibinfo {howpublished} {The unfolding procedure refers to
  the rescaling of the energies, so that the local density of states of the
  renormalized eigenvalues is 1} \BibitemShut {NoStop}%
\bibitem [{not({\natexlab{c}})}]{noteAve}%
  \BibitemOpen
  \href@noop {} {}\bibinfo {howpublished} {We also average the border defects
  over 20 random numbers to decrease the fluctuations after saturation and we
  do a running average} \BibitemShut {NoStop}%
\end{thebibliography}

%

\end{document}